\documentclass[twocolumn,showpacs,showkeys,preprintnumbers,amsmath,amssymb]{revtex4}

\usepackage{graphicx}
\usepackage{dcolumn}
\usepackage{bm}

\begin{document}
\preprint{preprint}

\title{The origin of the work function}
\author{Hennie C. Mastwijk}
\email{hennie.mastwijk@wur.nl}
\author{Paul V. Bartels}
\author{Huub L.M. Lelieveld}%
\affiliation{%
Wageningen University and Research,\\ 
Bornsesteeg 59, 6708PD, Wageningen, The Netherlands\\}%

\date{Submitted to APS 11-5-2006, revisions 14-12-2006, 5-3-2007, 31-5-2007, ....}%

\begin{abstract}
In this paper we analyse the mechanisms responsible for the bonding of electrons to metal surfaces. We present and validate a method to measure the energy distribution of dense electron ensembles at ambient conditions. The method relies on the detection of the density of states near the Fermi level. We have found sharp structures in the energy distribution of charge confined at the surfaces of Zn, Cu, Al, Si and C. These structures correspond to the onset of space charge i.e. electron emission. The threshold for emission is at an energy level that is associated to the work function $\phi$ of the metal under consideration. The work function increases linear with the free electron density and equals the exchange energy given by the Hartree-Fock model of electrons. We conclude that electron-electron interaction is the major factor that contributes to the observed work function of metals. In our experiments, electron emissions are triggered by the photoelectric effect. Remarkably, the emissions that have been observed do not obey the Einstein-Millikan threshold relation $h\ nu$=$\phi$ of the photoelectric effect. By discrimination between requirements for energy and momentum we deduce that the emissions are trigged by photon momentum. We claim that the photoelectric effect is the result of a phase transition in an electron gas that connects a condensed surface state to a free continuum state. The assignment of a phase transition to the effect under consideration is justified by analysing the behaviour of the Fermi-Dirac energy distribution upon raising the Fermi level. When the Fermi level crosses the  continuum level we predict an abrupt increment in the number of continuum state electrons that exceeds 70 orders of magnitude. Both the observed threshold energy and the line shape of the transition are in good agreement with theoretical predictions. 

\end{abstract}
\pacs{0375.Ss, 0530.Fk}%
\maketitle

\section{\label{sec:introduction}Introduction}
Quantum electronics has been an important research topic in industry and academia for the last century. Recent work includes studies on devices for cold electron emission using nano structures, studies on photonic devices and (bio-) chemical sensors [1-9]. Major efforts have been put to understand the specific role of surface states [9-14]. More fundamental issues concerning quantum many-body theory are subject to advanced experimental research using dense particle systems that are confined in optical lattices [15-17]. In the latter studies ions, atoms and molecules have been used to mimic degenerate electron gases. This approach allows the full control over many features of interest e.g. lattice parameters or differences caused by the symmetry of the wave function of ensembles, which accounts for the statistical behaviour of particles. This kind of experiments has been used for fundamental tests on the interaction between fermions, bosons and mixtures. Ultimately, this will lead to a full understanding of a long-standing problem known in solid state physics: the legitimacy of the independent electron approximation [18].

Whereas the majority of the experiments have been carried out under vacuum conditions at low temperatures [1-17] we will study dense electron systems at ambient conditions. Electron densities commonly found in metals yield $\sim 10^{23}cm^{-3}$. The corresponding deBroglie wavelength of electrons near the Fermi level exceeds the mean electron separation (Table I), even at a temperature of 300K. Thus, at ambient conditions the electron system in metals has to be treated as a quantum system in the low temperature limit [20, 21]. In this regime of high density and low temperature one has to go beyond the independent electron approximation and account for electron-electron interactions [18,21]. 

We consider the Hartree-Fock theory of the free electron gas [18] as a theoretical framework for the evaluation of exchange mechanisms. Although the many electron wave function of an ensemble is treated by a single particle wave function this theory accounts for electron-electron interactions. The electron-electron interaction gives rise to the so-called exchange energy term that is given as a function in the electron density ($1/r_s$) by [18]

\begin{equation}
\epsilon_{exch}=-\frac{3}{16\pi^2}\frac{q^2 k_f}{\epsilon_0}=-0.456\frac{q^2}{4\pi \epsilon_o r_s}\
\label{eq:exchange}
\end{equation}
                      
This term represents the average binding energy of a single electron to the ensemble as a result of exchange interaction. Note that the attractive coulomb interaction [21,22] of the proton-electron system (hydrogen) is identical to the exchange interaction of the electron-electron system, except for an (exact) factor of $(\frac{15}{16 \pi^2})^{1/3}$=0.4562720. 

Evidently, the exchange mechanism is the origin of an attractive coulomb force between electrons, which is responsible for the cohesion of dense electron ensembles. In Fig. 1 we have plotted the exchange energy as a function of the electron density for the metal elements. We have made a comparison with the work function of these elements known from literature [18,29].

From the many observations that have been made e.g. by photon spectroscopy and thermo ionic emission [18,19] it is apparent that electrons are bound to a metal by some kind of process. However, the origin of the bonding process is not well understood [18,19]. The work function has been originally introduced as a phenomenological quantity to account for the bonding of electrons to metals. It is generally assumed that ion-electron interactions at the metal-air interface play an important role in the binding process [18]. Based on the analysis presented in Fig. 1 it is concluded that the work function is correlated to exchange energy i.e. we suggest that the binding is more electron-electron like. This is similar to the original work of N.D. Lang and W. Kohn [22] at the start of the formulation of density functional theory [18].  

\begin{figure}
\centerline{\includegraphics[width=60mm,angle=-90,clip]{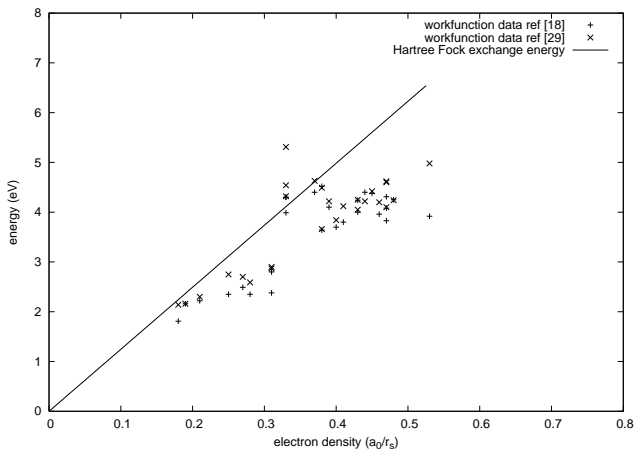}}
\caption{\label{fig:work function} The work function (data points) and the Hartee-Fock estimate of exchange energy due to electron-electron interaction (solid line) for metals with different electron densities.}
\end{figure}

\begin{table*}
\caption{\label{tab:table1}Electronic properties of different metals relevant to this work [18].}
\begin{ruledtabular}
\begin{tabular}{ccccccc}
Element&Density&Mean      &Fermi&deBroglie&\multicolumn{2}{c}{Work} \\
       &       &separation&level&wavelength&\multicolumn{2}{c}{function (eV)} \\
			 &$(\times 10^{22}cm^{-3})$& ($r_s/a_o)$&(eV)&(nm)&[29]& [18]\\
\hline
Be	&	24.7	&	1.87	&	14.3	&	0.31	&	4.98	&	3.92	\\
Al	&	18.1	&	2.07	&	11.7	&	0.34	&	4.25	&	4.25	\\
Fe	&	17	&	2.12	&	11.1	&	0.35	&	4.6	&	4.31	\\
Mn	&	16.5	&	2.14	&	10.9	&	0.35	&	4.1	&	3.83	\\
Sb	&	16.5	&	2.14	&	10.9	&	0.35	&	4.63	&	4.08	\\
Ga	&	15.4	&	2.19	&	10.4	&	0.36	&	4.2	&	3.96	\\
Sn	&	14.8	&	2.22	&	10.2	&	0.36	&	4.42	&	4.38	\\
Bi	&	14.1	&	2.25	&	9.9	&	0.37	&	4.22	&	4.4	\\
Zn	&	13.2	&	2.3	&	9.47	&	0.38	&	4.05	&	4.24	\\
Pb	&	13.2	&	2.3	&	9.47	&	0.38	&	4.25	&	4	\\
In	&	11.5	&	2.41	&	8.63	&	0.4	&	4.12	&	3.8	\\
Tl	&	10.5	&	2.48	&	8.15	&	0.41	&	3.84	&	3.7	\\
Cd	&	9.27	&	2.59	&	7.47	&	0.43	&	4.22	&	4.1	\\
Hg	&	8.65	&	2.65	&	7.13	&	0.44	&	4.49	&	4.52	\\
Mg	&	8.61	&	2.66	&	7.08	&	0.44	&	3.66	&	3.64	\\
Cu	&	8.47	&	2.67	&	7.0	&	0.44	&	4.64	&	4.4	\\
Au	&	5.9	&	3.01	&	5.53	&	0.49	&	5.31	&	4.3	\\
Ag	&	5.86	&	3.02	&	5.49	&	0.5	&	4.54	&	4.3	\\
Nb	&	5.56	&	3.07	&	5.32	&	0.5	&	4.33	&	3.99	\\
Li	&	4.7	&	3.25	&	4.74	&	0.53	&	2.9	&	2.38	\\
Ca	&	4.61	&	3.27	&	4.69	&	0.54	&	2.87	&	2.8	\\
Sr	&	3.55	&	3.57	&	3.93	&	0.59	&	2.59	&	2.35	\\
Ba	&	3.15	&	3.71	&	3.64	&	0.61	&	2.7	&	2.49	\\
Na	&	2.65	&	3.93	&	3.24	&	0.65	&	2.75	&	2.35	\\
K	&	1.4	&	4.86	&	2.12	&	0.8	&	2.3	&	2.22	\\
Rb	&	1.15	&	5.2	&	1.85	&	0.86	&	2.16	&	2.16	\\
Cs	&	0.91	&	5.62	&	1.59	&	0.92	&	2.14	&	1.81	\\
n-Si	&	5.2	&	3.14	&	-	&	-	&	-	&	-	\\
C (graphite)	&	$\sim 3\ 10^{18} cm^{-3}$	&	81	&	-	&	-	&	-	&	-	\\
\end{tabular}
\end{ruledtabular}
\end{table*}

However, in this paper we will investigate this hypothesis more carefully using statistical physics and experimental proofing rather than attempting to understand the physics by {\it ab inito} calculations as usually explored in quantum chemistry. In our experiments we shall use the dense electron ensembles that are already present at the surface of common metals. We shall consider the valence electrons as free electrons near the Fermi level. In a next step, we will introduce a method to measure the density of states of an electron ensemble at the Fermi level. By observing the structure in the density of states of an ensemble as a function of the number of electrons present we shall be able to quantify the contribution of exchange energy to the work function.

\section{\label{sec:theory}Theoretical description}

In our experiments we will consider surface and space charge and treat them as free electron gases. We will control the Fermi level of ensembles by changing the total number of particles that is contained in the gas. Therefore, we will first evaluate the statistical aspects of a localised ensemble that contains a variable amount of electrons. The electrons present in a gas obey Fermi-Dirac statistics and the energy distribution is given by [18-20]

\begin{equation}
FD(E,V)=\frac{1}{e^{(E-E_f)/kT}+1}
\label{fermi dirac}
\end{equation}

for $E > E_f$ and where $E_f$ denotes the Fermi energy of the ensemble. The continuum energy level is located at $E=0$. The total number of electrons that is present is determined by the external static electric potential of the ensemble which is denoted by $V$. The Fermi level is given by $E_f=E_{fo}-qV$ where $E_{fo}$ is the Fermi level in absence of this external potential. At a negative potential the Fermi energy is increased as a result of the addition of charge to the system. When electrons are added, the Fermi level rises to the value $E_{fo}-qV$. This is required because the number of electrons in the ensemble has increased whereas the maximum occupation number of successive energy states is limited to two (Pauli Exclusion Principle). Thus, the Fermi level of an electron ensemble is shifted in presence of an external potential by an amount of $-qV$ whereas the ground state level is fixed.

Now, we introduce a potential well to represent a confining structure (e.g. a crystal). As usual [18,20], we define a constant potential of magnitude $-\phi-E_{fo}$ in the region where the crystal is present and zero elsewhere. By introducing this potential both the ground state level and the Fermi level in the crystal are shifted by an amount $-\phi-E_{fo}$ [18-20]. Both levels are shifted by the introduction of a confining potential but the number of electrons did not change. This means that the Fermi energy has not changed. Thus, the Fermi level is now located at $-\phi-qV$ and the ground state level at $-\phi-E_{fo}$.

By choice of these definitions, electrons are in the continuum whenever $E>0$. Secondly, the work function is defined by $\phi$. The work function corresponds to the minimum energy that is required to promote an electron originally at the Fermi level of a charge neutral metal into the continuum [18-20]. 

The energy levels of the idealised electron system under consideration are schematically depicted in Fig. 2. This is the basic model that is used throughout this paper to clarify how the energy distribution of an electron ensemble is affected by charging a metal. In Fig. 4 we have evaluated the energy distribution of an electron ensemble held by a Zn plate ($E_{fo}=9.5eV, \phi=4.2 eV$) for the case that the plate is grounded, and for the case that it is charged to $-5V$. 

When a negative potential is applied the charge density has increased. For applied voltages less than $-4.2V$ the Fermi level exceeds the continuum level. All of the electrons that are added beyond this point are free from an energetic point of view. These electrons should, in principle, be released from surface into the continuum. In our experiments we will verify that this is actually the case by charging metals plates and probe for the emission of electrons. Before doing so, we have to consider the density of states of surface charge and relate the changes to the macroscopically observable charge and voltage.

\begin{figure}
\centerline{\includegraphics[width=60mm, angle=-90,clip]{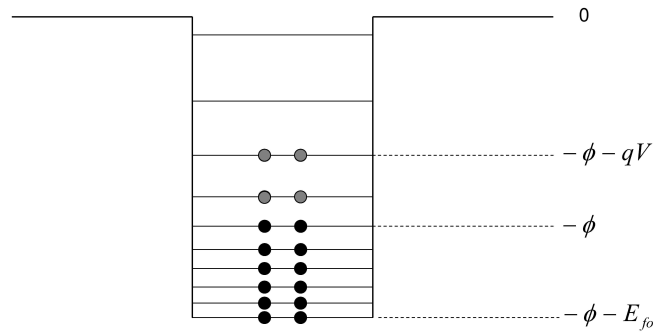}}
\caption{\label{fig:diagram}Relevant energy levels of a degenerate electron system confined in a potential well. The number of electrons is varied by means of an external potential. In absence of such a potential the levels are filled over a range of energies corresponding to the Fermi energy $E_{fo}$. The Fermi level is located at $-\phi$. Electrons are added to the system by charging the system to a negative potential $V$. After charging the Fermi level has risen to the level $-\phi-qV$.}
\end{figure}

\begin{figure}
\centerline{\includegraphics[width=60mm,angle=-90,clip]{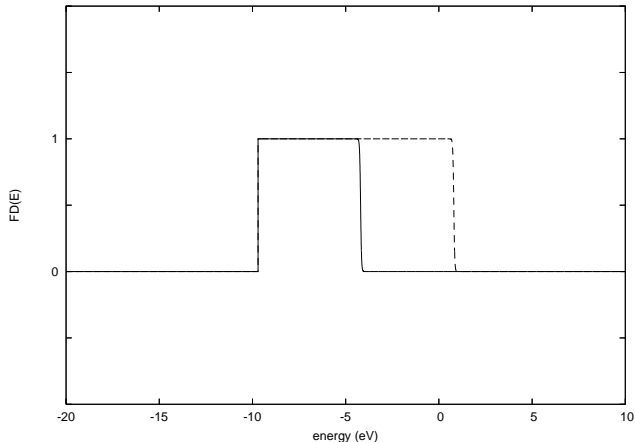}}
\caption{\label{fig:Fermi-Dirac evaporation} The energy distribution of electrons at a Zn surface with $\phi$=4.2eV and $E_{fo}$=9.7eV at a bias potential of $0V$ (solid line) and for  $-5V$ (dashed line). Beyond $-4.2V$ the Fermi level is extended into the continuum (Fermi-Dirac evaporation).}
\end{figure}

We argue that the correct quantum mechanical description of surface charge is given by an ensemble of free, independent electrons with two degrees of freedom.  In classical electrostatic theory it is well established that negative charge at a conductor is located at the outer surface [23]. This is a direct consequence of Gauss' law that implies that in absence of an internal electric field the surfaces of conductors are equipotentials. The energy distribution of surface charge is given by the definition of capacitance $C$ through $Q=CV$ [22]. As the electric charge is quantized by the electron charge $-q$ we can write $Nq=-2CV$ for a macroscopic amount of charge that contains $N$ electrons. A factor of two is added to account for two spin states. If we use the definition of the Volt, which is the energy per unit of charge given by $E=-qV$, and differentiate, we obtain the energy relation for the classical surface density of electrons which is given by 

\begin{equation}
\frac{dn_c}{dE}=\frac{2C}{A q^2}.
\label{}
\end{equation}

Here $A$ denotes the surface area under consideration. In addition, it is well known that the capacity of metal objects is merely independent of the voltage if the electrical field strength just outside the object is limited to 20 kV/cm (in order to prevent electrical breakdown in air). To summarize: we have identified the capacity as a measure of the classical surface density of electrons i.e. the number of electrons per unit of energy. The quantity is independent of energy. 

As a quantum mechanical description of surface charge we consider a free electron gas with two degrees of freedom. In general, the single particle density is given by the expectation value divided by quantisation volume $\Omega$ 

\begin{equation}
\rho=\frac{1}{\Omega} \left| \Psi \right|^2 =\frac{1}{\Omega}.
\label{rho}
\end{equation}

For a free electron that is confined at a square surface of dimensions $L\times L$, the charge density $\sigma$ is given by 

\begin{equation}
\sigma=\frac{q}{L^2} \left| \Psi \right|^2=\frac{q}{L^2}. 
\label{sigma}
\end{equation}

The electron energy is given by [18-21]

\begin{equation}
E_{ij}=\frac{\hbar^2 \pi^2} {2 m L^2} [i^2+j^2].
\label{surface electron energy}
\end{equation}

For an electron in the ground state (i=1, j=1), $L^2$ can be eliminated by combining Eq. \ref{sigma} and \ref{surface electron energy} to obtain the density of states (dos) in terms of surface charge density

\begin{equation}
\frac{dn_{qm}}{dE}=\frac{1}{q} \frac{d\sigma}{dE}=\frac{m}{\pi^2\hbar^2}.
\label{dos square}
\end{equation}

This result was obtained after quantisation of the free particle wavefunction on a square. However, even on a flat surface the coulomb force is a central force. Thus, in principle the quantisation procedure should be carried out on a disk. This involves the introduction of Bessel equations and its solutions. Its introduction leads to mathematical complexities and does not lead to more insight in the underlying physics. Alternatively, we adopt the direct counting procedure of electron states as employed in standard textbooks on quantum physics [18-21]. This counting procedure is well established and involves a quantisation procedure that is effectively carried out in spherical coordinate system. The surface density of states that is obtained in this approximation is given by

\begin{equation}
\frac{dn_{qm}}{dE}=\frac{m}{\pi\hbar^2}.
\label{dos disk}
\end{equation}

The expectation value of the particle density of a surface quantum gas is independent of its energy similar as in the classical description of surface charge. Upon comparison of the classical result with the quantum mechanical result we can define a minimum length scale where quantum mechanical effects are to be anticipated. The macroscopic electron surface density obtained by Gauss' law can be expressed as 

\begin{equation}
\frac{dn_c}{dE}=\frac{2C}{A q^2}=\frac{2\epsilon_o}{q^2r}
\label{classical surface charge}
\end{equation}

where we have used $C=4\pi\epsilon_or$ for a classical sphere of radius $r$. When we demand that this is equal to the density of states of the quantum mechanical electron gas we obtain an expression of the radius of a charged sphere that carries the charge of a single electron. This radius is given by $r=2\pi\epsilon_o\hbar^2/mq^2$ which is equal to half of Bohrs' radius for hydrogen. For charge densities larger than $2q/a_o$ only a quantum mechanical description of the electron is appropriate. For charge densities smaller than $2q/a_o$ classical electrostatic theory is expected to hold. Note that this is consistent with the estimate of exchange energy in the Hartree-Fock model: exchange effects are only relevant at densities in the order of $q/a_{o}$ and vanish at low density. 

Apparently, a classical plane surface relates to a topological structure that does not contain microscopic structures smaller than $a_o$. Thus for classical electrostatic theory to hold, a metal surface should be free of quantum dots (zero Degrees Of Freedom, DOF), quantum wires (one DOF) or quantum wells (two DOF). This is consistent with the fact that the density of states of latter structures is dependent on energy [18,24]. Its presence leads to distinct structures in the observed density of states (often referred to as quantum capacitance) as measured by e.g. conductance spectroscopy [24]. 

That the equivalency of Eq. 8 and 9 makes sense from a physical point of view can be demonstrated more generally by using the virial theorem [21]. This theorem states that the average kinetic electron energy of a free electron is twice the average action of the (repulsive) coulomb force:

\begin{equation}
2E_k=\int_{r'>r}{F(r')r'dr'}=\frac{q^2}{4 \pi \epsilon_o r}=\frac{\hbar^2 k^2}{m}
\label{eq:virial theorem}
\end{equation}

\noindent which is implicated by setting Eq. 8 equal to Eq. 9. Note that the path integral of the action only contributes in the outer region of a charged sphere: the contribution over its interior is zero. From this result we deduce that a 'free' electron is always under influence of its own repulsive coulomb force [25] and that the (positive) coulomb energy is associated to its kinetic energy.   

The eigen energies of a quantum gas of independent surface electrons are given by Eq. \ref{surface electron energy}. In Table II we have explicitly given the lowest lying single particle energy levels. The levels are enumerated according to the momentum eigen states $p_{ij}=\hbar k(i,j)$ with $k=\pi/a$ and where $a$ is the quantisation length of the problem under consideration. In Fig. 4 we have plotted the density of states as a function of the electron energy for a gas that contains $10^4$ electrons. We have taken $a=1.0\ 10^{-7}m$ and have assumed an energy resolution of 25 meV. This corresponds respectively to the typical surface roughness of the metal surfaces and to the typical energy resolution in the experiments that are discussed in this paper. From Table II and Fig. 4 we learn that the charge density per unit of energy is a constant, as it should. However, this is only true when we consider the average of a sufficient number of states. 

Fluctuations originate from changes in the level of degeneracy of successive energy levels (see Table II). It is interesting to note that these fluctuations are directly related to the mathematical problem known in number theory on the representation of a natural number as the sum of two squares. 

Based on the discussion in the previous paragraphs we summarize that surface charge as treated by classical electrostatic theory is the limiting case of a free, quantum electron gas with two degrees of freedom. The classical limit corresponds to the quantum mechanical result in the limit of large quantisation lengths (or equivalent in the limit of $\hbar\rightarrow 0$) as required by Bohrs' correspondence principle [21]. We shall confirm experimentally that the charge density is independent of energy and that the characteristic fluctuations shown in Fig. 4 are real.

\begin{figure}
\centerline{\includegraphics[width=60mm, angle=-90,clip]{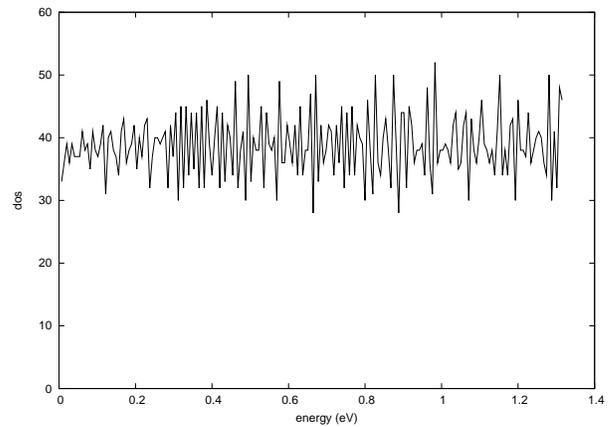}}
\caption{\label{fig:fluctuations} Calculation of the charge density as a function of energy for a two dimensional free electron gas that contains 10,000 electrons.}
\end{figure}

\begin{table}
\caption{\label{tab:table2}Lowest lying energy eigen states of an isotropic free particle quantum gas with two degrees of freedom. The momentum eigen states are given by $p_{ij}=\hbar k(i,j)$ at energy $E_{ij}=\frac{\hbar^2 k^2}{2m}\left[i^2+j^2\right]$.}
\begin{ruledtabular}
\begin{tabular}{cccccccc}
\# & $i$ & $j$ & $i^2+j^2$  & \#(cont) & $i$ & $j$ & $i^2+j^2$ \\
\hline
0	&	0	&	0	&	0	&	36	&	6	&	1	&	37	\\
1	&	0	&	1	&	1	&	37	&	2	&	6	&	40	\\
2	&	1	&	0	&	1	&	38	&	6	&	2	&	40	\\
3	&	1	&	1	&	2	&	39	&	4	&	5	&	41	\\
4	&	0	&	2	&	4	&	40	&	5	&	4	&	41	\\
5	&	2	&	0	&	4	&	41	&	3	&	6	&	45	\\
6	&	1	&	2	&	5	&	42	&	6	&	3	&	45	\\
7	&	2	&	1	&	5	&	43	&	0	&	7	&	49	\\
8	&	2	&	2	&	8	&	44	&	7	&	0	&	49	\\
9	&	0	&	3	&	9	&	45	&	1	&	7	&	50	\\
10	&	3	&	0	&	9	&	46	&	5	&	5	&	50	\\
11	&	1	&	3	&	10	&	47	&	7	&	1	&	50	\\
12	&	3	&	1	&	10	&	48	&	4	&	6	&	52	\\
13	&	2	&	3	&	13	&	49	&	6	&	4	&	52	\\
14	&	3	&	2	&	13	&	51	&	7	&	2	&	53	\\
15	&	0	&	4	&	16	&	52	&	3	&	7	&	58	\\
16	&	4	&	0	&	16	&	53	&	7	&	3	&	58	\\
17	&	1	&	4	&	17	&	54	&	5	&	6	&	61	\\
18	&	4	&	1	&	17	&	55	&	6	&	5	&	61	\\
19	&	3	&	3	&	18	&	56	&	0	&	8	&	64	\\
20	&	2	&	4	&	20	&	57	&	8	&	0	&	64	\\
21	&	4	&	2	&	20	&	58	&	1	&	8	&	65	\\
22	&	0	&	5	&	25	&	59	&	4	&	7	&	65	\\
23	&	3	&	4	&	25	&	60	&	7	&	4	&	65	\\
24	&	4	&	3	&	25	&	61	&	8	&	1	&	65	\\
25	&	5	&	0	&	25	&	62	&	2	&	8	&	68	\\
26	&	1	&	5	&	26	&	63	&	8	&	2	&	68	\\
27	&	5	&	1	&	26	&	64	&	6	&	6	&	72	\\
28	&	2	&	5	&	29	&	65	&	3	&	8	&	73	\\
29	&	5	&	2	&	29	&	66	&	8	&	3	&	73	\\
30	&	4	&	4	&	32	&	67	&	5	&	7	&	74	\\
31	&	3	&	5	&	34	&	68	&	7	&	5	&	74	\\
32	&	5	&	3	&	34	&	69	&	4	&	8	&	80	\\
33	&	0	&	6	&	36	&	70	&	8	&	4	&	80	\\
34	&	6	&	0	&	36	&	71	&	0	&	9	&	81	\\
35	&	1	&	6	&	37	&	72	&	0	&	9	&	81	\\
\end{tabular}
\end{ruledtabular}
\end{table}

\section{\label{sec:experiment}Experimental set-up}
We use polycrystalline plates of Zn, Cu, Al and C (as graphite) with dimensions of 40x40x2mm and a n-Si(100) plate as an electrode of an air capacitor. An aluminium grid (70$\%$ open, 1 mm mesh) forms the counter electrode. We study the transient current and voltage characteristics of this system by means of a discharge through a resistor. We use the photoelectric effect to liberate electrons from the surface. Emissions are observed at nearly field free (dV/dx$<$10V/cm) conditions. Experiments are carried out under ambient conditions. That is, at a pressure in the range of 950-1050 mbar, at temperatures in the range of 20-22$^o$C and in an open-air environment. The electrodes are continuously flushed with nitrogen ($\sim$ 2 L/min) to prevent oxidation. The experimental set-up is shown in Fig. 5. 

We use a cold cathode UV-C source (Philips UK TUV PL-S 9W) that emits up to 253 nm or a 150W quartz Hg lamp that emits up to 180 nm. The lamp spectrum of the 150W Hg lamp is given in Fig. 6, which was measured using an Avaspec 2048-2 spectrometer supplied with a UA-178-1100 grating. Both lamps are operated by standard lamp gear at a line frequency of 50 Hz. The light intensity at the surface of the metal plate is characterized by the light intensity in the 200-400 nm wavelength region and was measured by a SG01S SiC UV photo diode (Cree Research, US). The UV intensity at the surface is typically 24mW/cm$^2$ (150W lamp) and 5.6mW/cm$^2$ for the PL-S lamp.

\begin{figure}
\centerline{\includegraphics[width=60mm, angle=-90,clip]{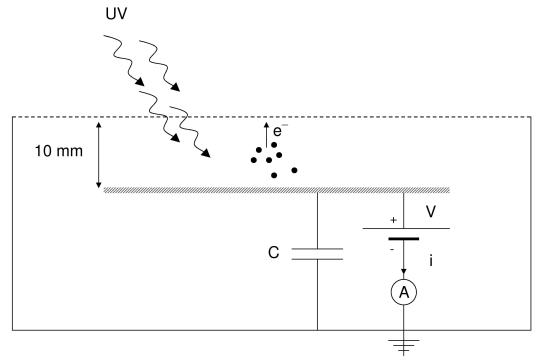}}
\caption{\label{fig:experiment}Schematic representation of the experimental set-up (side view). Coupons of Zn, Cu, Al, C or Si with dimensions of 40x40x2 mm are contained in a grounded shielding box. The top of the box is covered by an aluminium grid. Photons are introduced from the outside as depicted. A battery is used to provide a fixed bias voltage. Initially the metal plate is negatively charged by an auxiliary voltage source (not shown). Experiments are done at ambient conditions.}
\end{figure}

The voltage ($V$) and current ($I$) as indicated were measured using a Keithley 610C electrometer at an impedance setting of 100GOhm. Alternatively, the entire electro meter is replaced by a high impedance voltage divider (100:1) using two resistors (Ohmcraft HVRW 100GOhm and HVRW 1GOhm). By doing so we have verified that the voltage readings are not affected by biases that may originate from amplifier offset or unwanted contact potentials. Data are logged using a 16 bits ADC (Meilhaus FS 1608) at a rate of 6.6 samples per second. The accuracy of the voltage readings is 25mV and the absolute value is accurate within $\pm$100 mV. The potential of the metal plate under consideration is initially set using an auxiliary voltage source. Slowly varying voltages and currents are recorded when the system is allowed to discharge through the impedance of the electrometer.

\begin{figure}
\centerline{\includegraphics[width=60mm, angle=-90,clip]{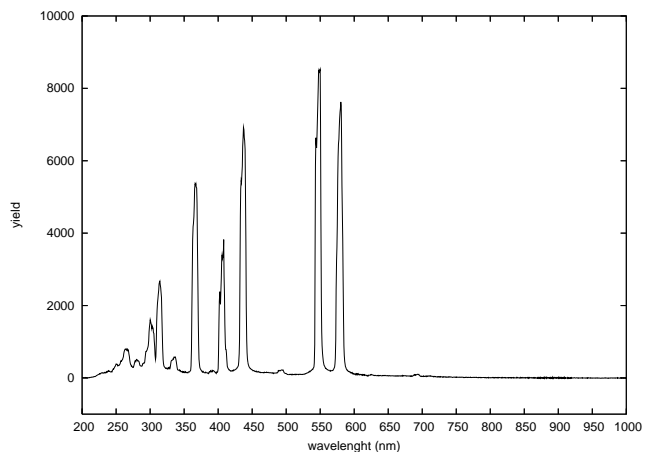}}
\caption{\label{fig:lampspectrum}Emission spectrum of the 150W quartz mercury lamp.}
\end{figure}

Oxide layers on the metal plates are removed using sanding paper (P400). The final surface roughness that is obtained after this procedure is 0.1-1.0 $\mu m$. By applying this procedure we obtain a randomly oriented polycrystalline surface. The silicon plate is a phosphorus doped Si wafer (n-Si, 5 $\Omega\ cm$, oriented in the (100) direction, Okmetic, Finland). It has an optical grade surface (surface roughness 1-3 nm) and is not scoured. 

In order to be able to measure the energy distribution over an energy range of -42eV to 50eV we initially charge the metal plate to -50V to hold an excess amount of electrons. The capacitor is then discharged by the RC circuit, either in the dark, or under illumination with UV light. A battery pack of +42V is used to provide the required positive bias and a 10nF capacitor is used to increase the amount of charge in the system that is initially present (Fig. 5). 

\section{\label{sec:analysis}Observations and analysis}

In absence of UV light the system discharges through the impedance of the voltmeter according to V(t)=V(0)$e^{-t/RC}$ with R=100G$\Omega$, C=10nF and V(0)=-92V. In presence of UV light, the capacitor is additionally discharged by a photoelectric current. Typical discharge curves that have been observed are plotted in Fig. 7. 

\begin{figure}
\centerline{\includegraphics[width=60mm,angle=-90,clip]{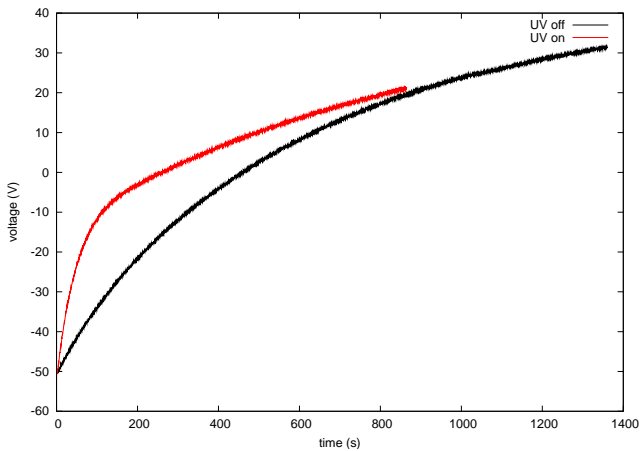}}
\caption{\label{fig:curves}Measured discharge currents in presence and absence of the UV light. The photoelectric effect is responsible for deviations in the exponential decay of the discharge current.}
\end{figure}

For a quantitative study on the underlying physical processes we study the density of states near the Fermi level. We therefore analyse and interpret our data in terms of changes in charge ($dQ$) and changes in energy ($dE$) rather than by current ($I$) and voltage ($V$). The conversion of the variables ($V,I$) into ($Q,dQ/dE$) is according to  $E=-qV$ and

\begin{equation}
\frac{dQ(E)}{dE}=-\frac{1}{q}\frac{I(t)dt}{V(t+dt)-V(t)}=-\frac{1}{q}\frac{I(t)dt}{dV(t)}.
\label{eq:transformation}
\end{equation}

This step is realised in our analysis by numerical differentiation. Note that the total charge per unit of energy $dQ(E)/dE \sim\ q dn(E)/dE$ is proportional to the density of states but requires the definition of a specific surface area or volume (see section V).

Our expression for the density of states is different from the one that is commonly used in scanning tunneling spectroscopy [24]. In the latter case, the local density of electron states is assumed to be proportional to $dI/dV$. we believe, however, that the density of states is a static parameter. The use of the differential conductance as a direct measure of the density of states is incorrect from a dimensional point of view. 

In Fig. 8.1a-8.5a we present the measured density of states of surface charge as a function of energy for Zn, Cu, Al, C and n-Si in absence of light. We have observed a constant level on average i.e. independent on energy. The fluctuations in the observed charge density have been attributed to the fluctuations also observed in Fig. 3. In our experimental set-up there are two contributions to these fluctuations. One from the metal foil contained in the 10nF capacitor and a second contribution is from the metal-air-aluminium grid capacitor ($\sim$ 1pF) under consideration. 

In our experiments we occasionally observe structures in the density of states at specific energy levels. We have found that these structures are the result of electronical surface processes (Redox reactions) at the metal-air interface. These structures appear as peaks upon a constant background and have a width of approximately 25meV. As oxidation reactions are known to occur at a specific energy and at a length scale on the order of $a_o$, these structures represent a quantum dot structure. Advanced diagnostics to observe these quantum dots and the dynamics of Redox reactions are topics of our current research and will be discussed elsewhere. At present we conclude that we have verified that Gauss' law applied to metal surfaces is an appropriate theoretical description of surface charge which corresponds to the quantum mechanical model of a two-dimensional free electron gas. 

In Fig. 8.1b-8.5b we have plotted the observed density of states of charge under the condition that the surface of the metal plate is illuminated by UV light. The intensity of the UV source at the surface of the plates yields 24mW/cm$^2$ in the case of Zn, Cu and Al and 96mW/cm$^2$ for C and n-Si. The spectrum of the UV source that has been used is given in Fig. 6. As a result of the emission of electrons by the photoelectric effect, we observe a decline in the occupation level of the surface states. This decline corresponds to an increase in the occupation level of continuum states.

The density of states of space charge is given by [18-21]

\begin{equation}
dn_{3d}(E)/dE=\frac{\sqrt{2}m^{3/2}}{\pi^2\hbar^3}\sqrt{E-E_o}. 
\label{space charge}
\end{equation}

The characteristic energy level $E_o$ relates to the threshold energy level at which electrons are considered being in the continuum. 

\begin{figure*}
\begin{tabular}{cc}
\vspace{-3mm}
(8.1a) \includegraphics[width=40mm, angle=-90,clip]{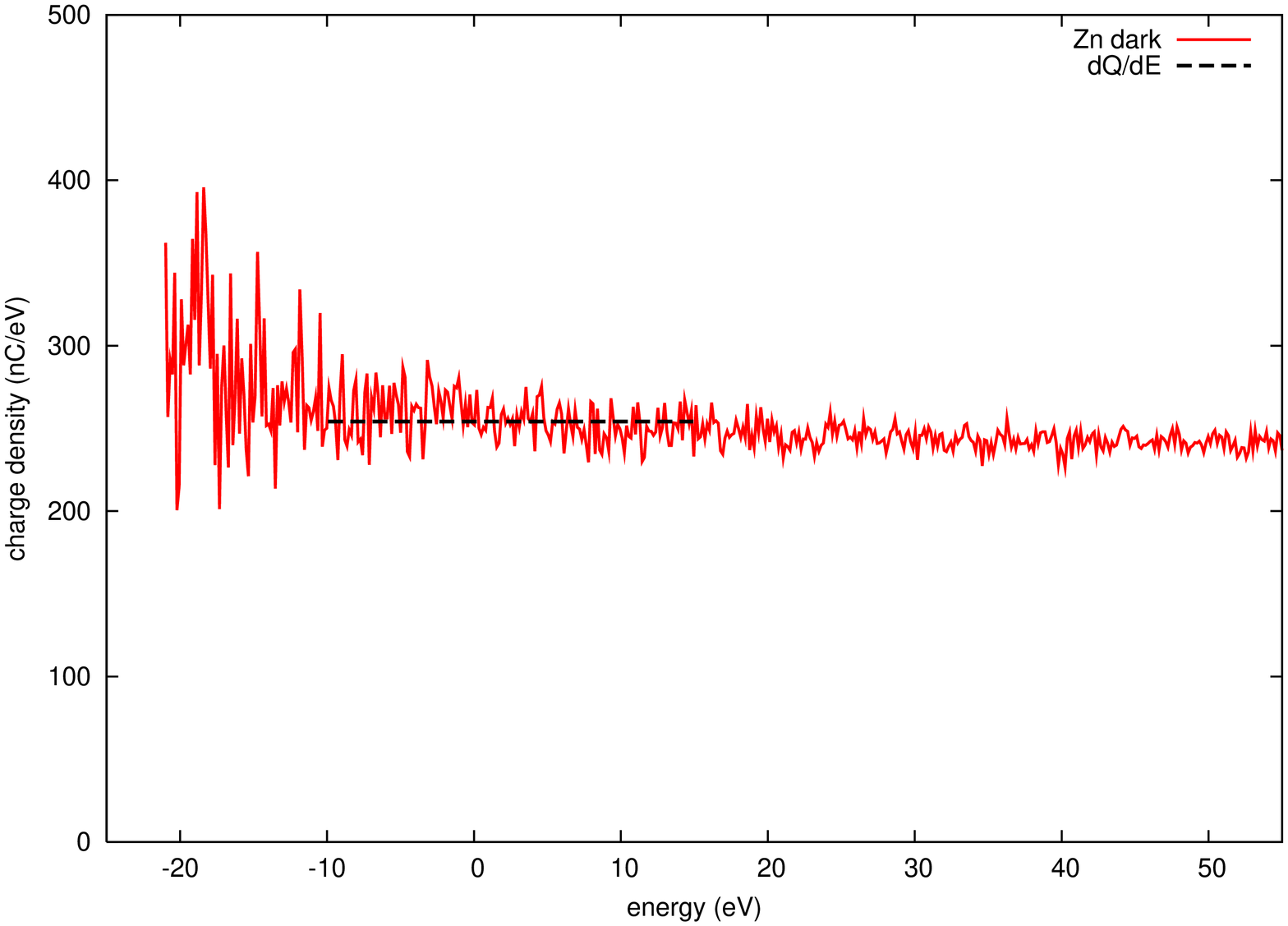}&
(8.1b) \includegraphics[width=40mm, angle=-90,clip]{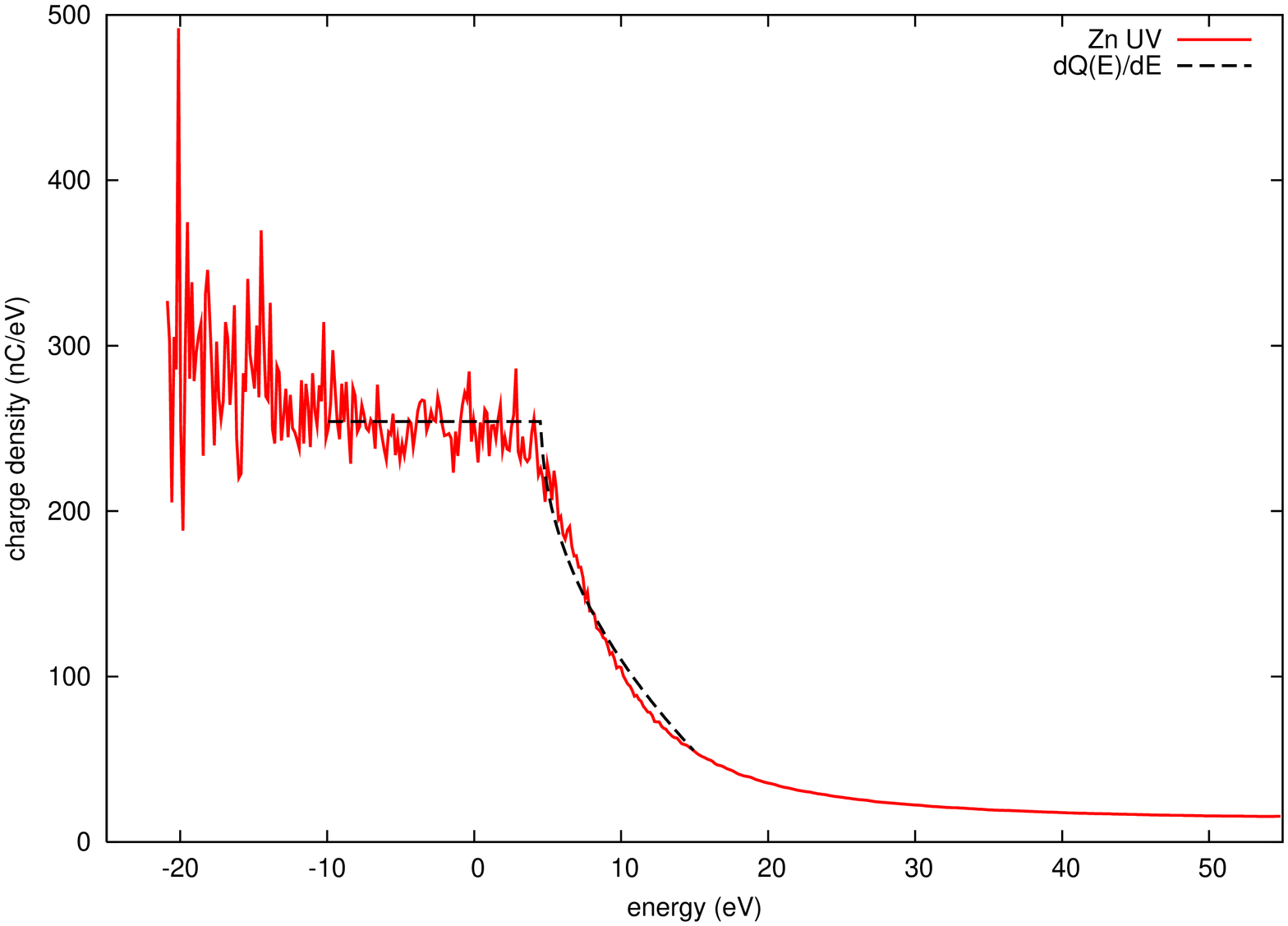}\\
\vspace{-3mm}
(8.2a) \includegraphics[width=40mm, angle=-90,clip]{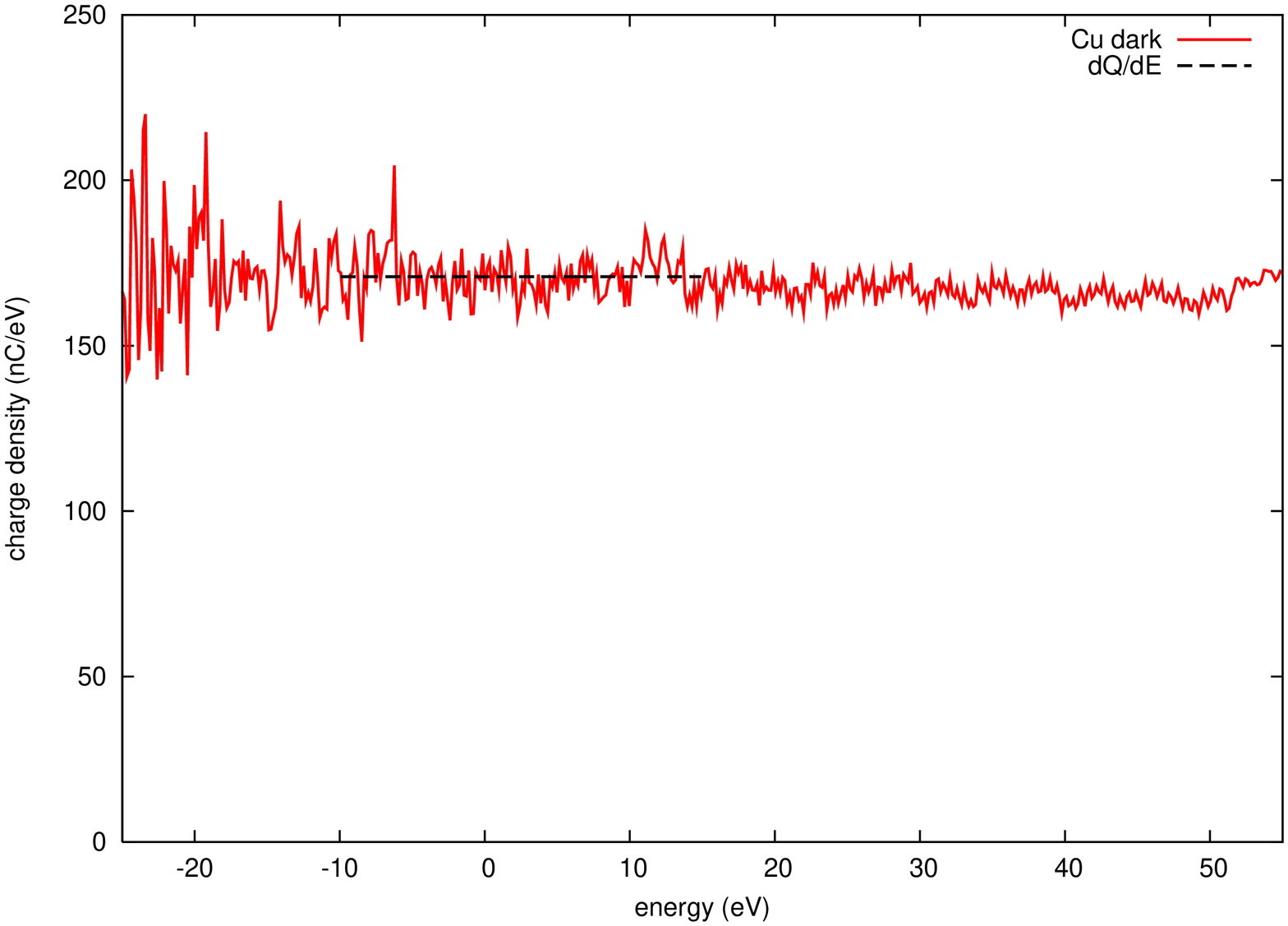}&
(8.2b) \includegraphics[width=40mm, angle=-90,clip]{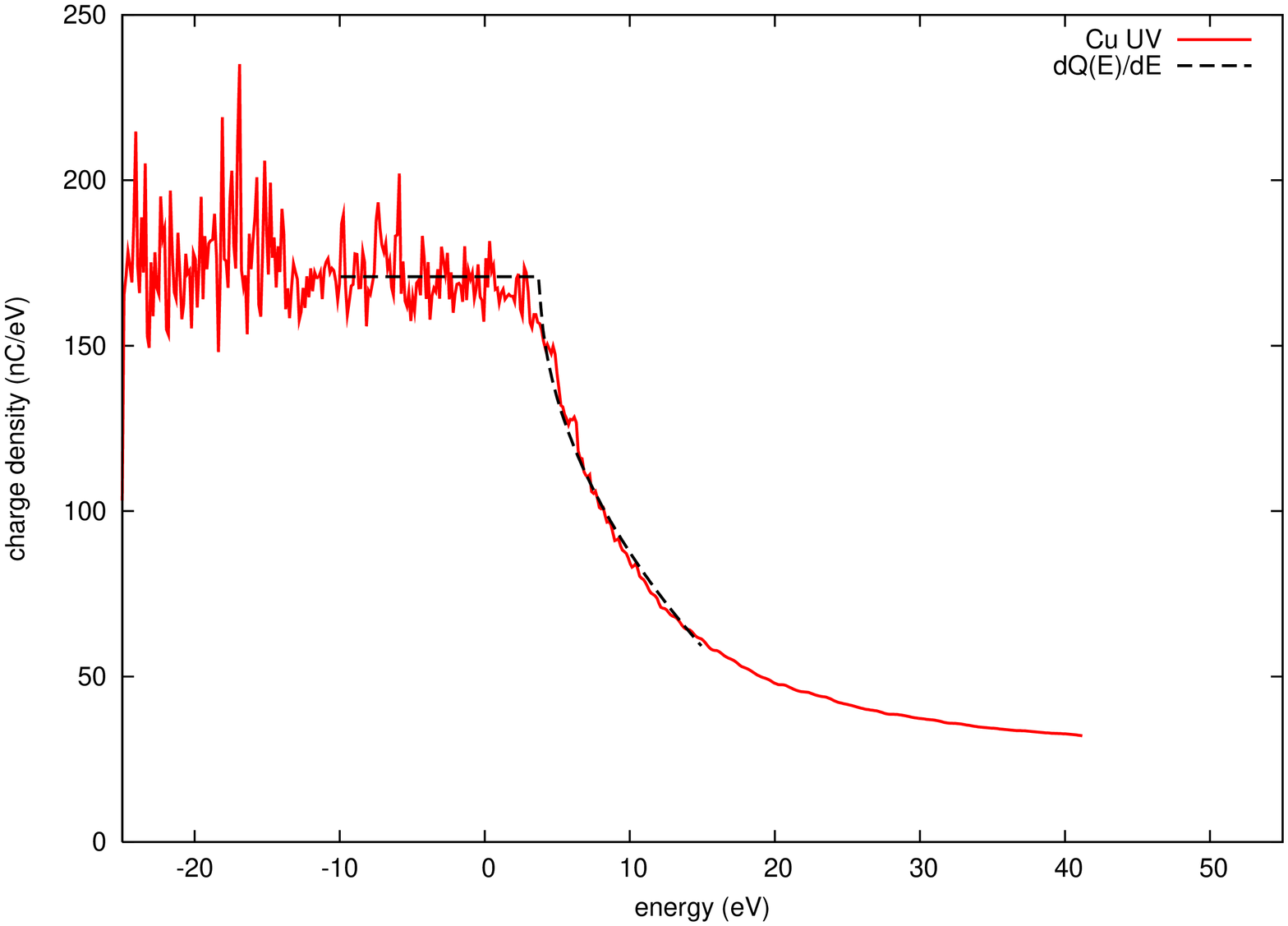}\\
\vspace{-3mm}
(8.3a) \includegraphics[width=40mm, angle=-90,clip]{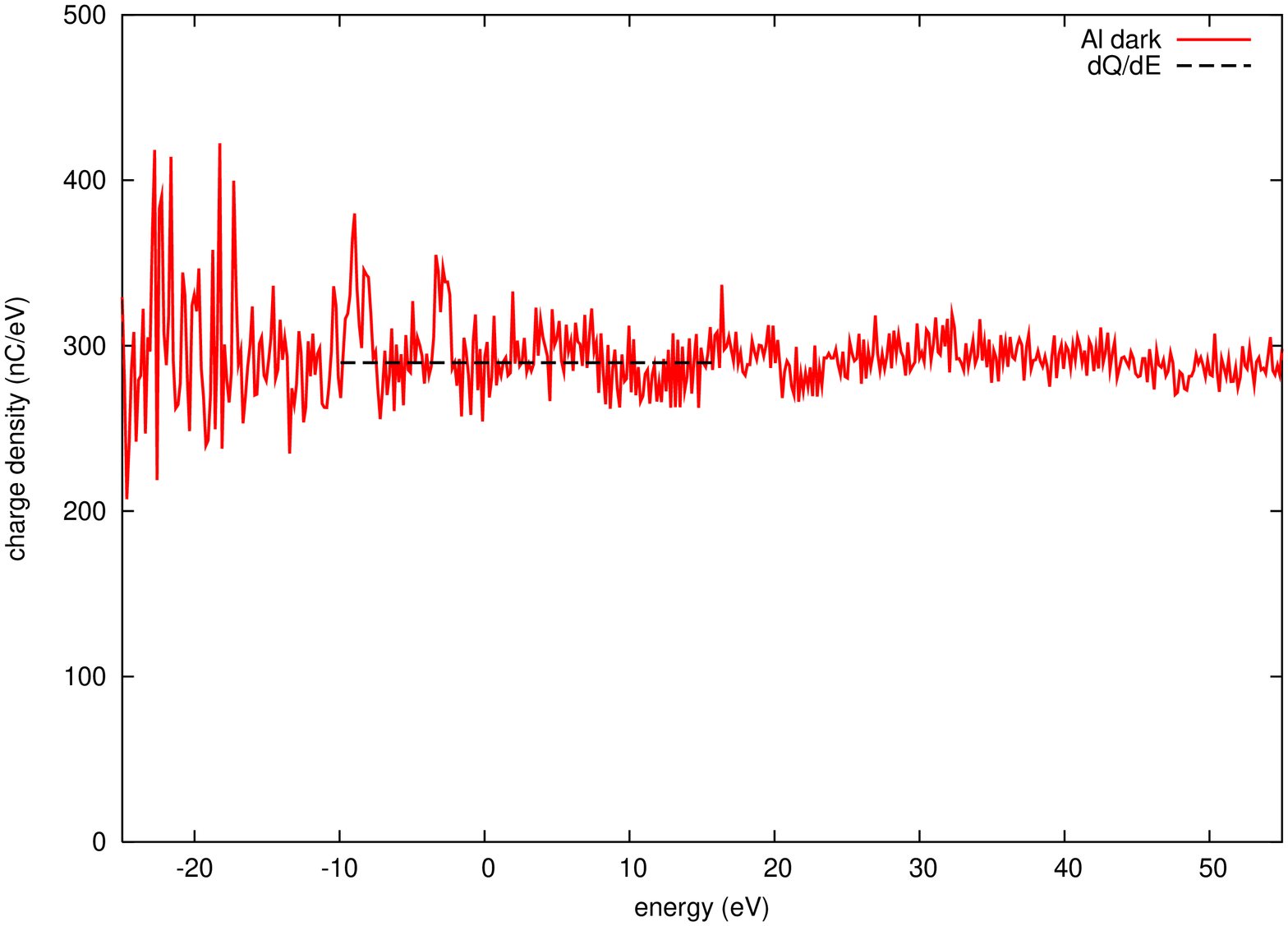}&
(8.3b) \includegraphics[width=40mm, angle=-90,clip]{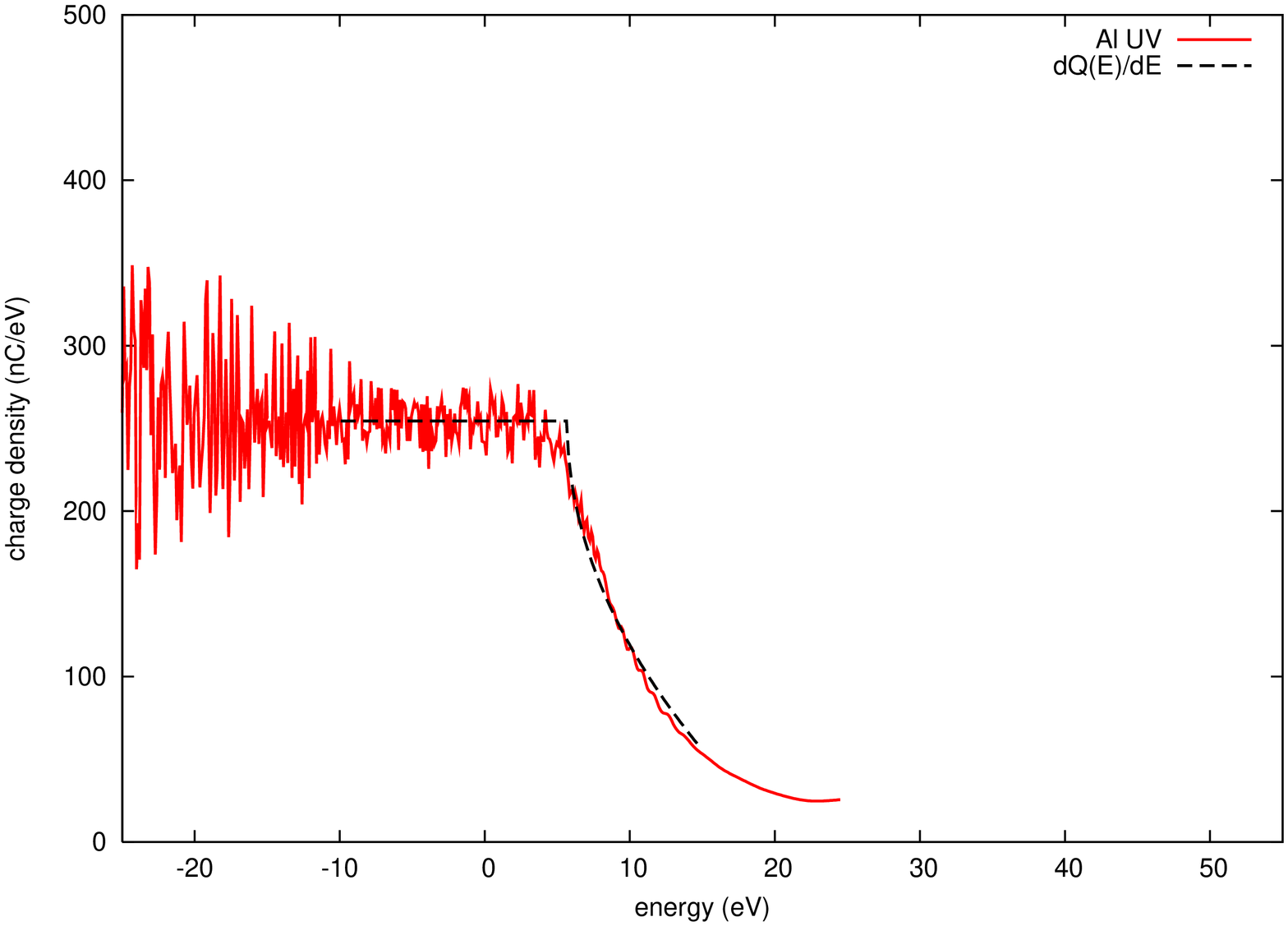}\\
\vspace{-3mm}
(8.4a) \includegraphics[width=40mm, angle=-90,clip]{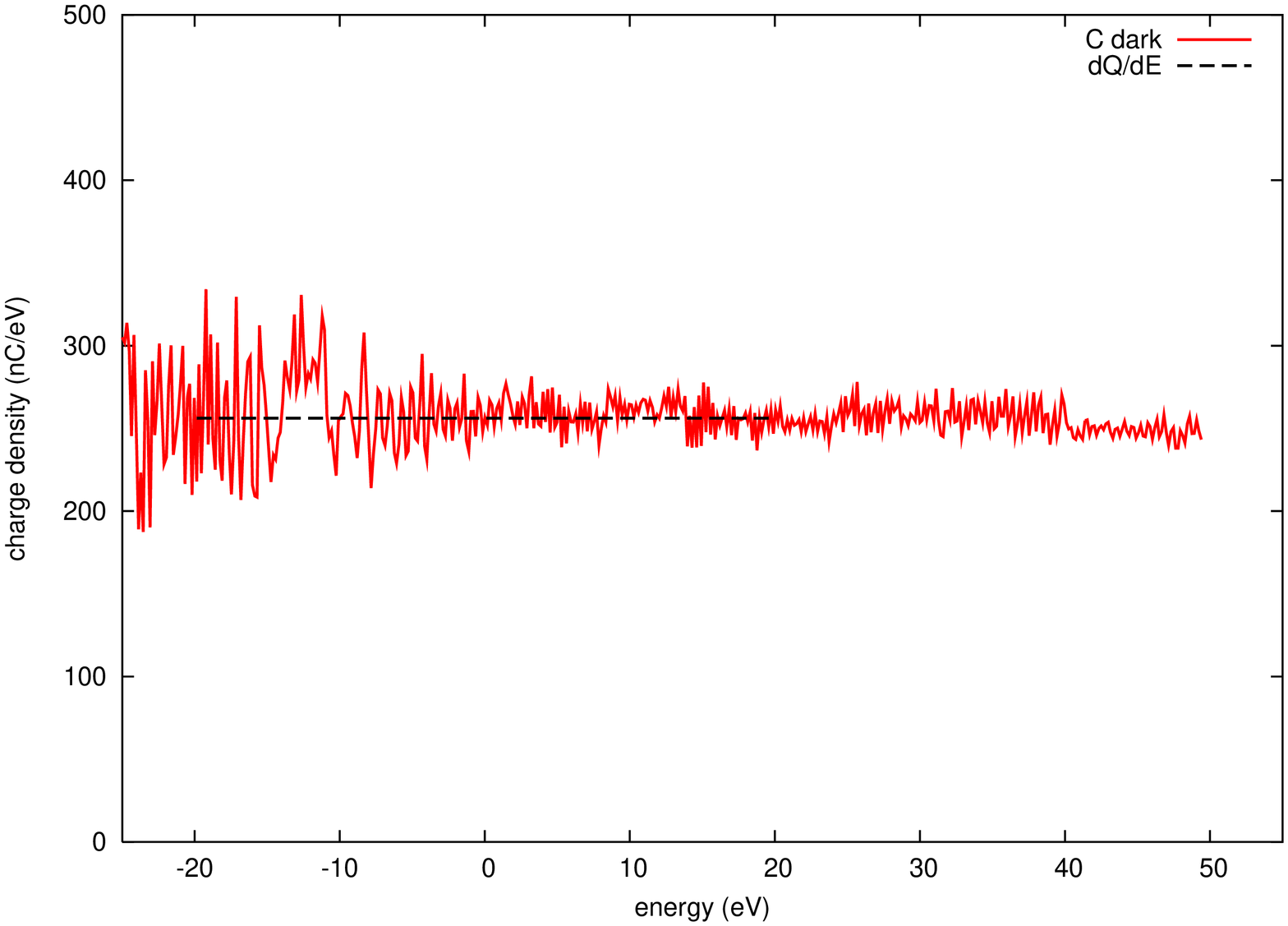}&
(8.4b) \includegraphics[width=40mm, angle=-90,clip]{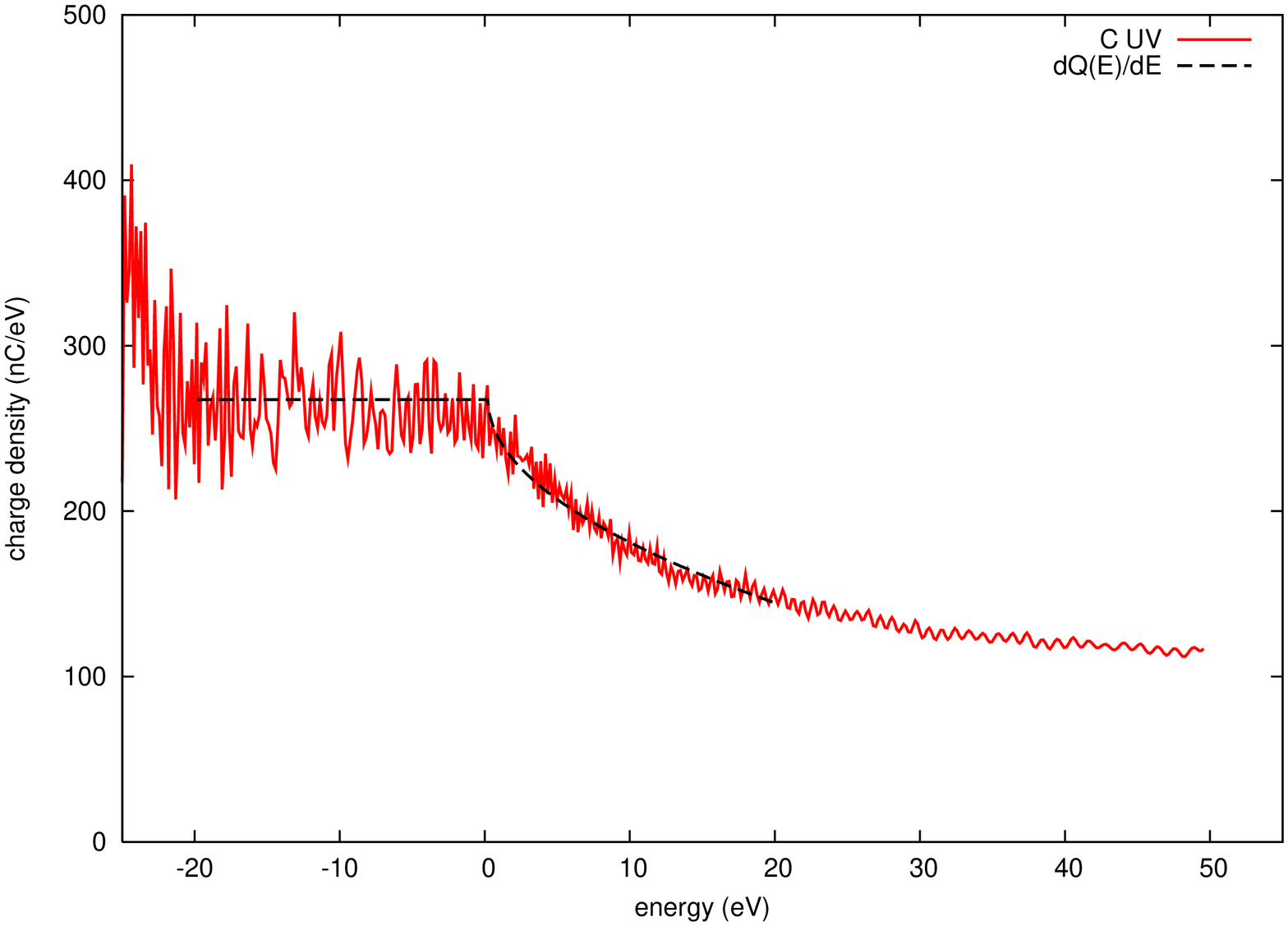}\\
\vspace{-3mm}
(8.5a) \includegraphics[width=40mm, angle=-90,clip]{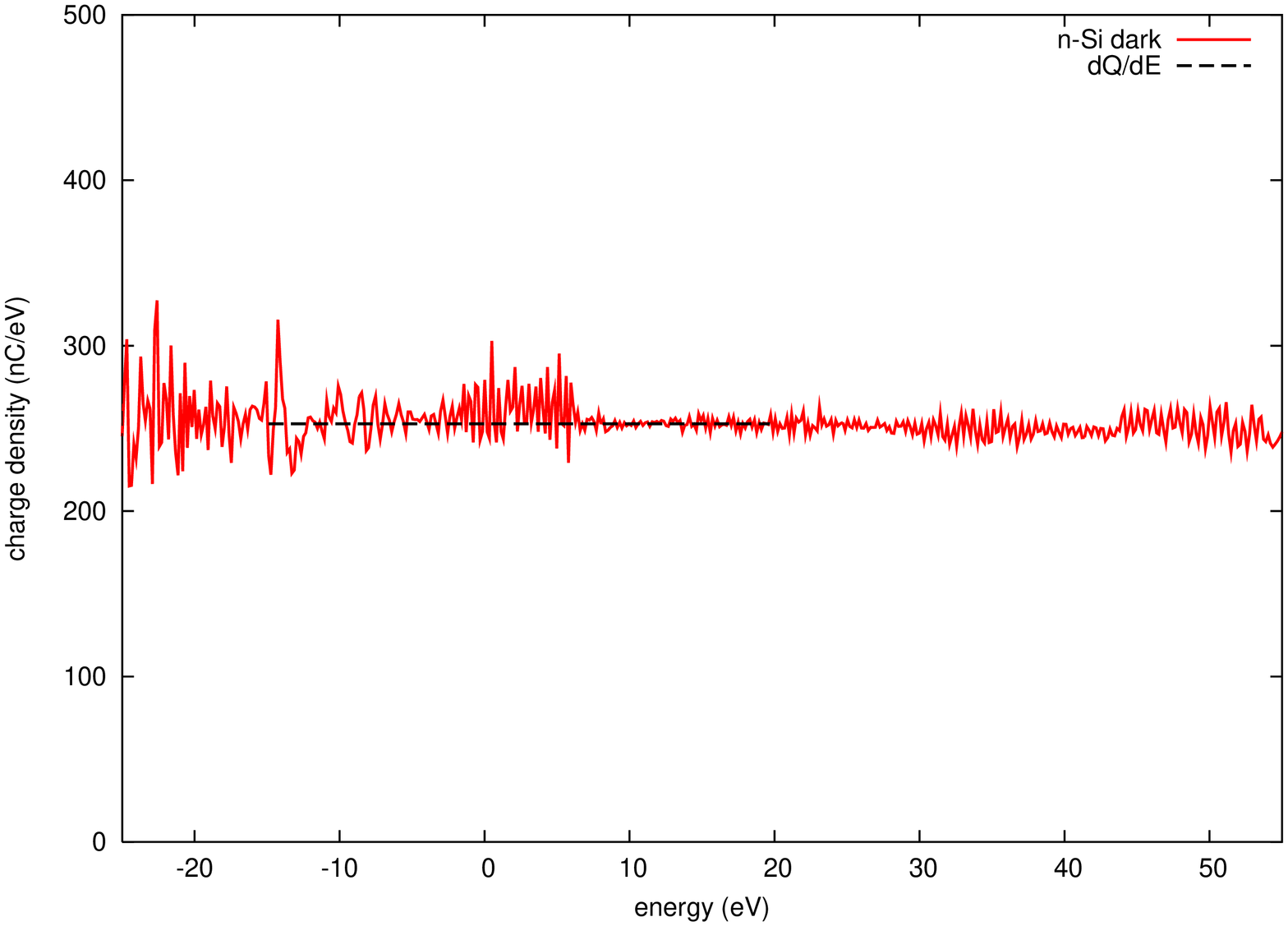}&
(8.5b) \includegraphics[width=40mm, angle=-90,clip]{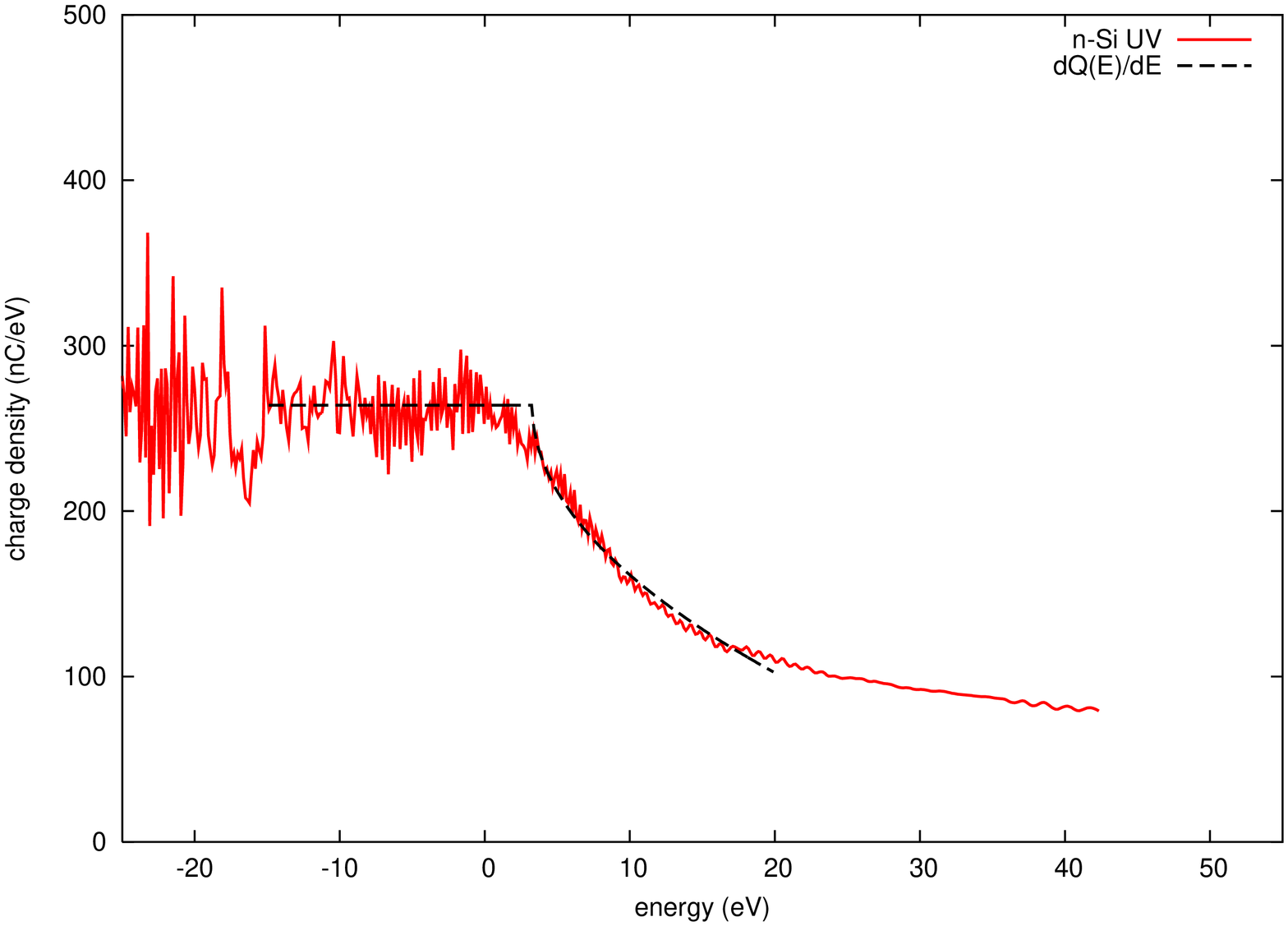}\\
 & \\
\end{tabular}
\caption{\label{fig:data}The charge density on metal plates as a function of the Fermi level. a) In absence of UV light the density of states is a constant. b) In presence of UV light the surface charge is emitted and the density of states is a combination of surface charge and space charge. The dashed lines are the fitted shapes according to Eq. 13. It represents the transition of a surface bound electron gas into the continuum.}
\end{figure*}

We have fitted the data sets in Fig. 8.1b-8.5b (dashed lines) to the energy distribution function of the injected charge $dQ/dE$ given by:

\begin{equation}
\frac{dQ(E)}{dE}=
\begin{cases} 
\frac{m}{\pi \hbar^2} A  & E<E_o\\
\frac{m}{\pi \hbar^2} A-\frac{\sqrt{2}m^{3/2}}{\pi^2 \hbar^3}\Omega \sqrt{E-E_o} & E>E_o
\end{cases}
\label{transition}
\end{equation}

where $A$ respectively $\Omega$ define an effective surface area and volume. We have used $E_o$ as a fitting parameter to obtain an estimate for the onset energy of space charge. We have fitted the line shape at the energy domain as indicated in Fig. 8.1b-8.5b. At higher energies ($>20 eV$) the temperature of the electrons in the continuum increases as a result of the ionisation of molecular nitrogen (data not shown). This results in an overestimate in the density of states in the high energy limit as predicted by Eq. 2 and 13. By using the fitting procedure as discussed we obtain a proper, unbiased fit of all our data to a single line shape. The threshold energies that have been determined are presented in Table III.

\begin{table}[t]
\caption{\label{tab:table3}Observed threshold energies for the photoelectric effect determined by the method that has been described in this work compared to the Hartree-Fock estimate for exchange energy [18].}
\begin{ruledtabular}
\begin{tabular}{cccc}
Element & $(a_o/r_s)$ & Observed threshold& Hartree-Fock\\
				&             &  energy (eV)      & Exchange energy (eV)\\
\hline
C & 0.01 &0.2$\pm$0.4 &0.1\\
n-Si &0.32 & 3.5$\pm$0.2&4.0\\
Cu & 0.37 & 3.8$\pm$0.2 & 4.6\\
Zn & 0.43 & 4.5$\pm$0.2 &5.3\\
Al &0.48 & 5.7$\pm$0.2&6.0\\
\hline
\end{tabular}
\end{ruledtabular}
\end{table}

\section{Discussion}

The observation of space charge, and more specific, the observation of an electron energy distribution that scales as $\sqrt{E}$, has been reported already in 1911 by C.D. Child [28]. Child studied the emission current density from hot filaments in vacuum tubes and concluded that the current is limited by the presence of space charge. The phenomenological description of this effect has led to the Langmuir-Child Law (or the three-and-a-half-power law). Upon differentiation of this law we obtain a density of states that is proportional to $\sqrt{E}$. No reports have been found in literature where an offset energy is observed. 

More recent work deals with cold emission processes of electrons held in nano structures [1-3]. The observed energy distribution found in our work is certainly different from electrons that result from high field field emissions [1-3,18,19,24]. In the latter case electrons are emitted at high electrical field strengths ranging from 10 to 50kV/cm. The energy distribution follows a Fowler-Nordheim law as a result of quantum tunneling through the cathode-anode barrier [1-3]. Emissions that have been observed in our work occur in the energy range of 0 to 5V, which is nearly two orders of magnitude less than reported in [1-3]. To ascertain that we actually deal with a different emission process, we studied the onset of field emission in our set-up. In order to observe electron currents resulting from this process we had to increase the electric field strength to a level in excess of 30 kV/cm. This was realised by reducing the distance between the metal plate and the grid to 5mm and switching to an operating voltage of 15kV (data not shown). 

We studied the onset energy for the photoelectric effect for a range of metals: Zn, Cu, Al, C and n-Si. In all cases we observed a positive threshold energy. Zn has a work function of 4.2 eV, a value that is accurate within 5\% [19]. We studied the onset energy for Zn by varying the UV photon energies from 6.7eV to 4.5eV. We found that the onset energy for emission at +4.5 eV does not change within the experimental error of $\pm$0.2eV. By analysis of the onset energy found for Zn, Cu and Al (Table III) we can conclude that emission occurs in general when the surface is charged and the electrons reach an energy level associated with the work function of the metal. The specific behaviour of the Fermi-Dirac distribution, which has been discussed in Fig. 3, accounts for a sharp increase in the amount of space charge when the Fermi level exceeds the continuum level. Thus, by statistical analysis alone, an accurate quantitative prediction has been made for the work function which is determined by the onset energy for space charge. 

The observed emissions cannot be explained by the conventional interpretation of the photoelectric effect. In the latter case, the threshold energy for the emission of photoelectrons for Zn is expected at $\phi-h\nu=-2.4eV$ when using 180nm (6.7eV) photons [26-27]. As a result, these observations let us no other choice than to conclude that the observed emissions are of different origin. This conclusion may be implausible from a historic point of view. However, the Einstein-Millikan threshold relation $h\nu=\phi$ accounts for the minimum photon energy based on the measurement of the kinetic energy of electrons upon arrival. This is different in our experiments, where we control the Fermi level of an electron ensemble and observe when emission occurs. With certainty, electrons that satisfy the Einstein-Millikan relation are also released in our experiments. However, these ejected electrons have energy of a few eV above the Fermi level. They therefore loose their kinetic energy in collisions with gas molecules that are present in the ambient environment leading to a recapture by the surface. As a result these electrons do not contribute to the process of electron emission. It is by the construction of the experiment that we study the fate of the electron system as a whole. This is in contrast with the stopping voltage method [26] which is designed to determine the photon energy by kinetic analysis of the photo electron. It should not come as a surprise that the physical processes involved are of different origin. At this stage we conclude that our results do not contradict the Einstein-Millikan relation.  

The Fermi level in this work is controlled by electrostatic charging. As a consequence, the emission of electrons by the photoelectric effect is not limited by energy requirements. Indeed, we have observed an onset in emission at the point where the Fermi level coincides with the continuum level. We conclude that nearly free electrons are involved in the emission process. This implicates that the ejection is the result of a binary process, involving one photon and one electron. We consider the simplest process: an elastic collision of a photon and an electron (Compton scattering). The process is schematically depicted in Fig. 9. As energy is conserved in the scattering process, the loss in the photon energy is equal to the kinetic energy $E_k$  that is gained by the electron. In a head-on collision ($\theta=\pi$) this yields $E_k=\hbar^2 k^2/2m$, where $k$ is the wave vector of the incident photon. For visible light, the energy that is exchanged in the process is on the order of 10$\mu$eV. This is insignificant when compared to the amount of thermal energy that is available at 300K, which yields kT=26 meV. The amount of momentum that is transferred to the electron is considerable. It equals twice the photon momentum i.e. $p=2\hbar k$. The average amount of momentum that is exchanged by the Compton process equals $p=\hbar k=2\pi\hbar/\lambda$. We argue that this corresponds to the minimum amount of momentum that is needed to escape from a surface.

\begin{figure}
\centerline{\includegraphics[width=30mm, angle=-90,clip]{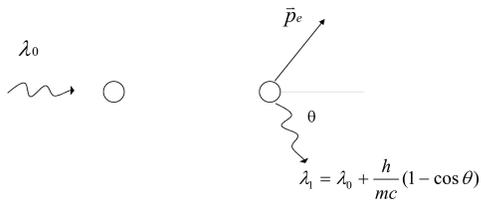}}
\caption{\label{fig:compton}
Elastic scattering of electrons by photons (Compton scattering). Left) A photon approaches an electron Right) The electron is scattered and has gained a momentum in the process. The energy and momentum exchanged is determined by the Compton shift $\lambda_1-\lambda_0$.}
\end{figure}

For an electron to escape a surface momentum is needed. This is the consequence of the broken translation symmetry near the metal-air interface, in compliance with Noethers' theorem on conservation laws and symmetry [21]. In order to satisfy the conservation of total momentum, a source is required that delivers an electron a sufficient amount of momentum at emission. We shall estimate the minimum amount that is needed by evaluation of the kinematic aspects of the ejection process. 

Consider the momentum eigen states $p=\hbar \pi/a(i,j,l)$ of an isotropic free electron gas, given by a set of quantum numbers $(i,j,l)$ that corresponds to the energy states $E_{ijl}=\hbar^2k^2/2m \left[ i^2+j^2+l^2 \right]$ where $k$ stands for the magnitude of the wave number associated to the electron. This wave number is determined by the quantisation length $a$ through $k=\pi/a$. For surface charge i.e. an electron gas with two degrees of freedom, $l$=0 and only $(i,j)$ states are occupied. For an electron in the continuum no restrictions have to be made and all $(i,j,l)$ states have to be considered. This means that in a transition from the surface into the continuum a minimum amount of momentum is required perpendicular to the interface of magnitude $p=\hbar k=h/a$. The quantisation length $a$ is specified by the local symmetry of space near the surface that is defined by the surface structure of the lattice at the metal-air interface. Structure is defined either by geometrical constraints or by the presence of impurities. In our experiments the surface structure of the metals and the graphite is specified by the microscopic surface roughness which lies in the range of 0.1-1 $\mu m$. From the dopant concentration in our n-Si sample ($\sim$ 10$^{15}$cm$^{-3}$) we deduce that the average distance between impurities yields $\sim$ 0.1 $\mu m$. Thus, the quantisation length of the surface structures in our experiments is in the order of the wavelength of UV light. We believe that the observed threshold wavelength of the light $\lambda_{thr}$, corresponds to the threshold of electron momentum $\pi/a$ that is needed to remove an electron from the surface: $\lambda_{thr}=2a$. This is in agreement with our observations that the threshold wavelength does not depend on the Fermi energy. For this, we have charged Zn plates to -100V, which is far beyond the continuum level and illuminated the surface with intense sources of light (UV-B, UV-A and visible light), all of well defined wavelength. We have probed for the onset of emission and found that the threshold wavelength to observe electron emission lies below 300nm. We therefore conclude that under the experimental conditions presented in this work the ejection of electrons is triggered by Compton scattering.

For carbon we have found a value for the work function that yields 0.2$\pm$0.4 eV (see Fig. 8.4b and Table III). This value is consistent with our assumption that exchange energy is the major contribution to the work function. Graphite is a semi-metal with a free electron density several orders of magnitude lower that in metals [18,19, Table I]. Exchange mechanisms are therefore absent, and as a result, the work function is non-existent. 

Silicon differs from the other elements under consideration since the sample that was used is crystalline. Due to the hardness of the material its surface could not prepared to obtain a polycrystalline sample. No attempts where made to remove any residual oxides from the surface. We have found structures in the surface density of states in the region of 0-5eV (Fig. 8.5a and 8.5b) which we attribute to band structure effects in the bulk and surface decontamination. Despite the fact that the silicon data in Fig. 8.5b deviates from the fitted model in the region of 0-5eV we have included the estimate for the threshold energy in Table III. This is motivated by the fact that a detailed, microscopic understanding of these effects is not a prerequisite for a correct description of the macroscopic parameters that describes a many particle system [20]. In this work we use statistical physics to describe the macroscopic parameters surface charge, Fermi energy and exchange energy. The use of statistical physics guarantees that the qualitative and the quantitative behaviour of a system is well described, at least in first order, independent of microscopic details. As our statistical model discribed by Eq. 13 fits to the overall shape of the Si data and since the threshold is a reasonable estimate of the work function assuming a bulk electron density for phosphorus doped n-Si, we have included Si in our work. 

Based on all of the results presented in Table III we conclude that the work function of a metal is linearly dependent on the electron density $1/r_s$ according to $W=(11.0 \pm 0.4) a_o/r_s$. This is consistent with the scattered work function data found in literature, presented in Fig. 1, which yields $W=(9.9) a_o/r_s$ on average. As the Hartree-Fock theory predicts that the exchange energy is a linear function in electron density we conclude that the bonding of electrons to a surface is dominated by the process of electron-electron interaction. The remaining discrepancy between the experimental values found and the Hartree-Fock estimate is attributed to screening. The effect of the screening of positive ions by strongly bound electrons is neglected by the model, but tends to reduce the exchange energy by 10-20\% [18].
 
By taking into account the sharp onset in the energy distribution that has been observed in Fig. 8.1b-8.5b, we suggest that the photoelectric effect is the result of phase transition from a condensed phase (surface charge) to a vapour (space charge). This is motivated by the observation that the transition can be crossed only in one direction (initially an excess amount of negative charge is needed). Secondly, we have found that the sharp onset coincides with the energy needed to break electron-electron bonds. This occurs when the Fermi level of an electron ensemble held at a surface, has been risen by an amount that matches the work function. It results in a phenomenal increase in the amount of space charge by a factor of $e^{qV/kT}=e^{162}$ which yields 70 orders of magnitude. This abrupt change in the occupation number of electrons in the continuum states (i.e. evaporation) is similar to the behaviour of the distribution function required for bosons in the low temperature limit. The number of bosons in the ground state increases abruptly (i.e. condensation) after a thermal equilibrium is reached below the Bose-Einstein transition temperature [20,21]. In the case of a dense electron system thermal equilibrium is only reached at temperatures that are in the order of the Fermi temperature. This associated temperature typically yields T=20,000 K for metals. At thermodynamical equilibrium with black body radiation at these extreme temperatures, the average photon energy of $\overline{E}=hc/\overline{\lambda}=2.3\ 10^{-4}\ T $=4.6eV and the photon density of $\overline{n}=3.0765/\overline{\lambda}^3\approx 10^{20}m^{-3}$ [20,21] are certainly sufficient to account for emissions. This corresponds to the high temperature limit where thermo-ionic emissions (T$\sim$ 2000 K) are observed [18,19]. In the low temperature limit black body radiation does no longer account for emissions. We found, however, that the presumed thermal equilibrium can be effectively mimicked by raising the Fermi level into the continuum and illumination of the surface by an auxiliary source of UV photons (Fermi-Dirac evaporation).

Finally, we argue that the photoelectric effect is an induced phase transition i.e. a process that is triggered after applying a force by using an analogy with superheated water. In the latter case, the temperature of the liquid phase exceeds the temperature of the vapor phase. Only after applying force to the water surface, the vanderWaals bonds between the water molecules are broken. This results in a sudden release of water vapor leading to the immediate cooling of the liquid. In this way the thermodynamical equilibrium between the liquid and vapor phase is restored. This is similar to surface charges held by a conductor at a negative potential; its excess charge is abruptly released after introducing photons which mediate forces on electrons.

\section{Conclusions} 
We have presented and validated a method to measure the energy distribution of dense electron ensembles at ambient conditions. The method relies on the direct measurement of the density of states at the Fermi level. By applying Bohrs' correspondence principle we have found that surface charge, as treated by Gauss' law in classical electrostatic theory, is the macroscopic limit of a free quantum electron gas with two degrees of freedom. The Fermi level charge held at the surface of a metal plate is determined by its electrical potential through $E=-qV$. When the Fermi level matches the work function, a sharp onset in the density of states is observed for Zn, Cu, Al, n-Si and C (as graphite). The structures correspond to the onset of space charge i.e. electron emission at respectively 4.5$\pm$0.2, 3.8$\pm$0.2, 5.7$\pm$0.2, 3.5$\pm$0.2 and 0.2$\pm$0.4 eV. The characteristic onset energy equals the work function. Remarkably, the work function of Carbon (a low electron density material) is vanishing. After comparison of the experimental values for the work function to the Hartree-Fock model of a free electron gas we conclude that exchange energy is the major factor that contributes to the work function of metals.  

Electron emissions were triggered in our work by the photoelectric effect. However, the observed emissions in this work do not obey the Einstein-Millikan threshold relation $h\nu=\phi$ of the photoelectric effect. By discrimination between requirements for energy and momentum, we find that the observed emissions in this work are triggered by photon momentum. Therefore we conclude that the photoelectric effect is the result of a phase transition in an electron gas that connects a surface state with two degrees of freedom to a continuum state with three degrees of freedom. The assignment of a phase transition to the observed effect has been justified by analyzing the behaviour of the Fermi-Dirac energy distribution. When the Fermi level is risen above the continuum level, the amount of space charge increases abruptly. The increase in the number of continuum electrons is phenomenal and exceeds 70 orders of magnitude. The experimental threshold energies and the line shapes of the transition are in good agreement with theoretical predictions.  

\begin{acknowledgments}
Support on electronics and radiometry were kindly provided by F. Ditewig, C.J. Wisman and J.G. Kornet ($\dagger$ 11-2-2007). The silicon wafers were provided by R. Luttge. We thank G. Ventura for providing the Ohmcraft specialty resistors and P. van der Straten and J.W. Thomsen for the fruitfull discussions and their critical comments. This work was financially supported by Senter-Novem.
\end{acknowledgments}

\newpage 

\end{document}